\documentclass[aps,prd,floatfix,showpacs,twocolumn,superscriptaddress]{revtex4-1}
\usepackage{amsmath}
\usepackage{amssymb}
\usepackage[dvipdfmx]{graphicx}
\usepackage{bmpsize} 
\usepackage[dvips]{color}
\usepackage[usenames,dvipsnames]{xcolor}
\newlength{\colw}
\newcommand{\One}{1\kern-4.5pt1}

\definecolor{DarkGreen}{rgb}{0,0.5,0}

\begin{document}

\title{Reply to "Comment on 'Lattice Gluon and Ghost Propagators, and the Strong Coupling in Pure SU(3) Yang-Mills Theory: Finite Lattice Spacing and Volume Effects' "}

\author{Anthony G. Duarte}
\author{Orlando Oliveira}
\author{Paulo J.\ Silva}
\affiliation{CFisUC, Department of Physics, University of Coimbra, P--3004 516 Coimbra, Portugal.}

\begin{abstract}
The quenched gluon and ghost propagator data published in~\cite{Duarte:2016iko} is reanalysed following the suggestion of~\cite{Boucaud:2017a}
to resolve the differences between the infrared data of the simulations.  Our results confirms that the procedure works well either for the
gluon or for the ghost propagator but not for both propagators simultaneously as the observed deviations in the data follow opposite patterns. 
Definitive conclusions require improving the determination of the (ratios) of lattice spacings.
A simple procedure for the relative calibration of the lattice spacing in lattice simulations is suggested.
\end{abstract}

         
\maketitle

The lattice studies of the gluon~\cite{Cucchieri:2007md,Cucchieri:2007zm,Bogolubsky:2009dc,Dudal:2010tf,Cucchieri:2011ig,Ilgenfritz:2010gu,Oliveira:2010xc,Oliveira:2012eh,Sternbeck:2012mf,Oliveira:2012eh,Duarte:2016iko}
and ghost~\cite{Sternbeck:2005vs,Oliveira:2006zg,Cucchieri:2007zm,Cucchieri:2008fc,Bogolubsky:2009dc,Ilgenfritz:2010gu,Cucchieri:2013nja,Cucchieri:2016jwg} propagators in pure Yang-Mills gauge theories has been thoroughly pursued in the past years. 
The emerging picture
being a finite and non-vanishing gluon propagator in the infrared region, a manifestation of a non-perturbative mechanism responsible 
for the generation of a 
gluon mass,
and a ghost propagator which follows closely its tree level value.

The production of high precision data for the propagators, module possible Gribov copies effects~\cite{Silva:2004bv}, 
requires understanding the finite volume $V$ and finite lattice spacing $a$ 
artefacts. 
In~\cite{Oliveira:2009eh,Oliveira:2012eh,Duarte:2016iko} a tentative was made to estimate the combined effects of using
a finite volume and a finite spacing on the simulations. For hypercubic lattices with a size $L a \gtrsim 6.5$ fm, in the
perturbative scaling window, the lattice propagators associated to simulations with a $\beta \gtrsim 5.7$  collapse into a single curve 
for momenta $p \gtrsim 1$ 
GeV. 
In the infrared region the lattice data differ by far more than one standard deviation,
revealing a systematic effect which 
remain to be understood.

\begin{figure}[h] 
   \centering
   \includegraphics[width=3in]{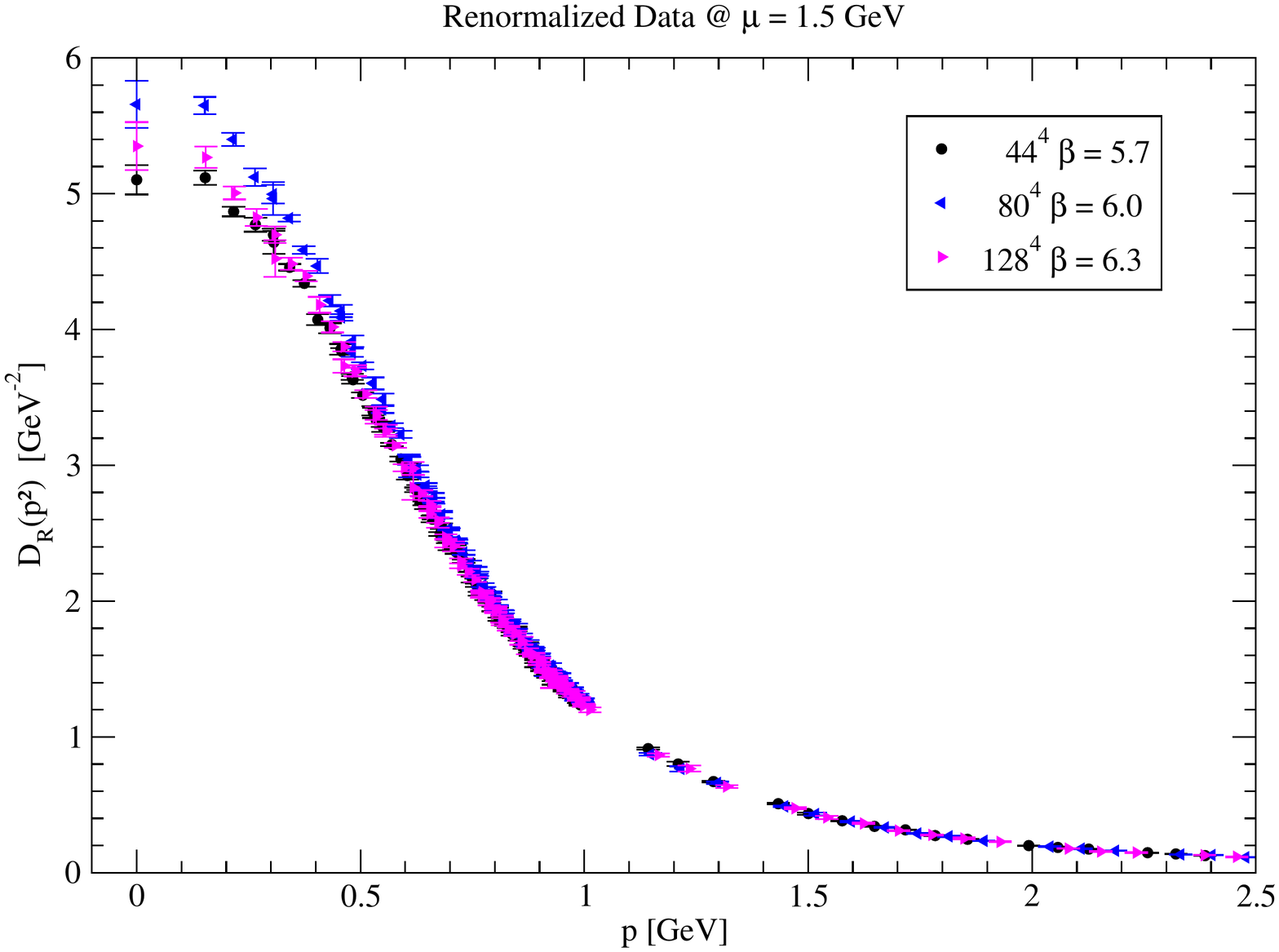} \\
   \includegraphics[width=3in]{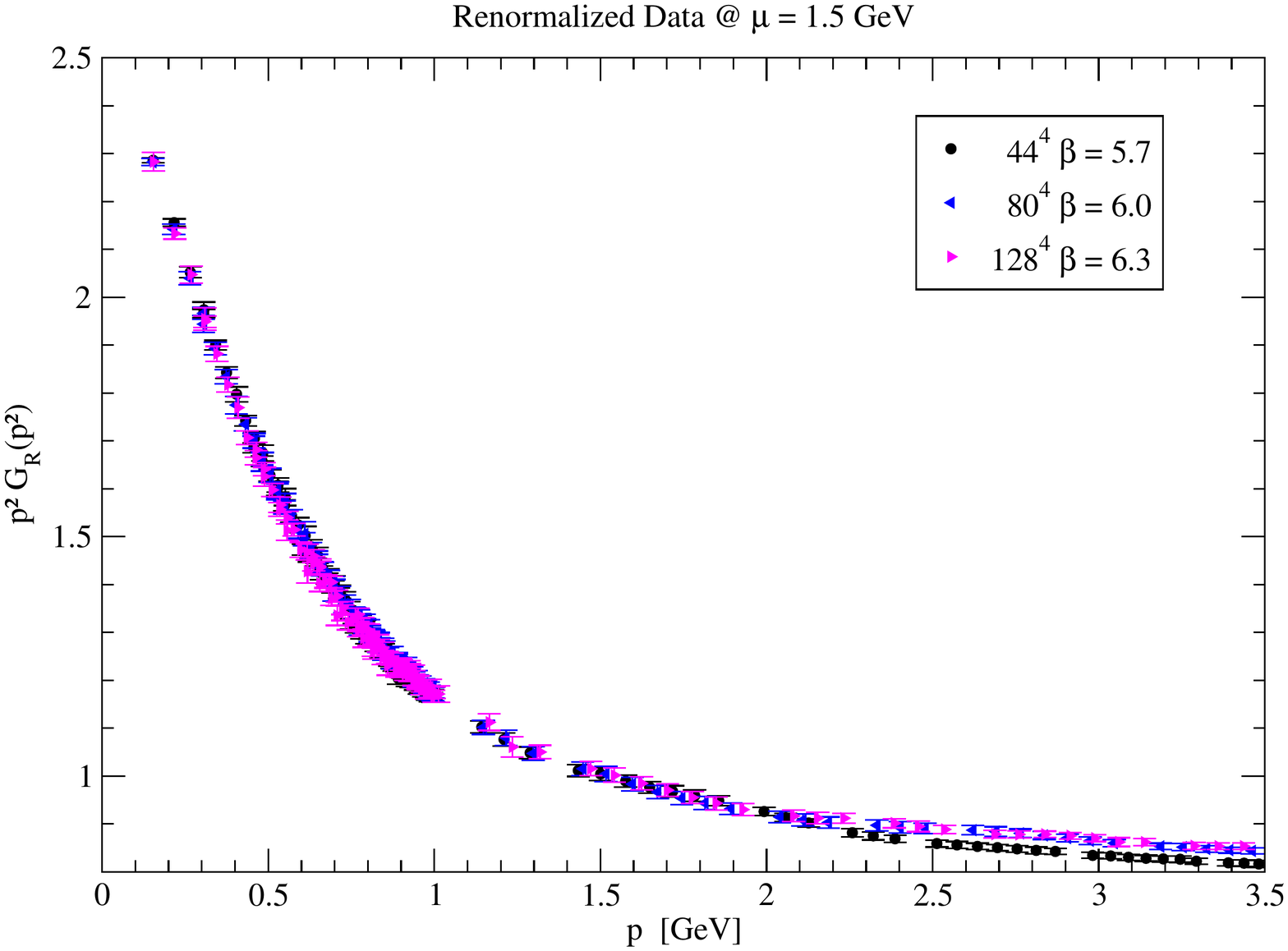}
   \caption{Renormalized gluon propagator (top) and ghost dressing function (bottom).} 
   \label{fig:props}
\end{figure}

In~\cite{Boucaud:2017a} the authors suggested that the observed differences can be 
attributed 
to the uncertainties in setting the scale in lattice simulations. 
Moreover, an example is given that by  ``\textit{recalibration}'' of the lattice spacing, compatible with the magnitude of the statistical
error on $a$, two different gluon data that were initially incompatible in the infrared region collapse into a unique curve. 

Despite the statistical error associated to any definition of the lattice spacing, the simulations for the propagators performed so far
never considered this effect on the final result. 
Note that this ``uncertainty'' 
is not related to lattice artefacts or to Gribov copies effects.
From Tab. I in~\cite{Duarte:2016iko} the lattice spacing reads $a = 0.1838(11)$ fm for $\beta = 5.7$, $a = 0.1016(25)$ fm for $\beta = 6.0$
and $a = 0.0627(24)$ fm for $\beta = 6.3$ which translates into a relative statistical error of $0.6\%$, $2.5\%$ and $3.8\%$.

The aim of  this reply is to redo the analysis of the data published in~\cite{Duarte:2016iko} for the gluon and ghost propagators
assuming the point of view of~\cite{Boucaud:2017a}. 
In order to avoid and reduce possible systematics due to the use of a finite lattice spacing, our first step is to renormalize
the data of~\cite{Duarte:2016iko} at a different kinematical point and we set
$D_R ( \mu ^2 ) = 1 / \mu^2$ with $\mu = 1.5$ GeV for both propagators. 

\begin{figure}[h] 
   \centering
   \includegraphics[width=3in]{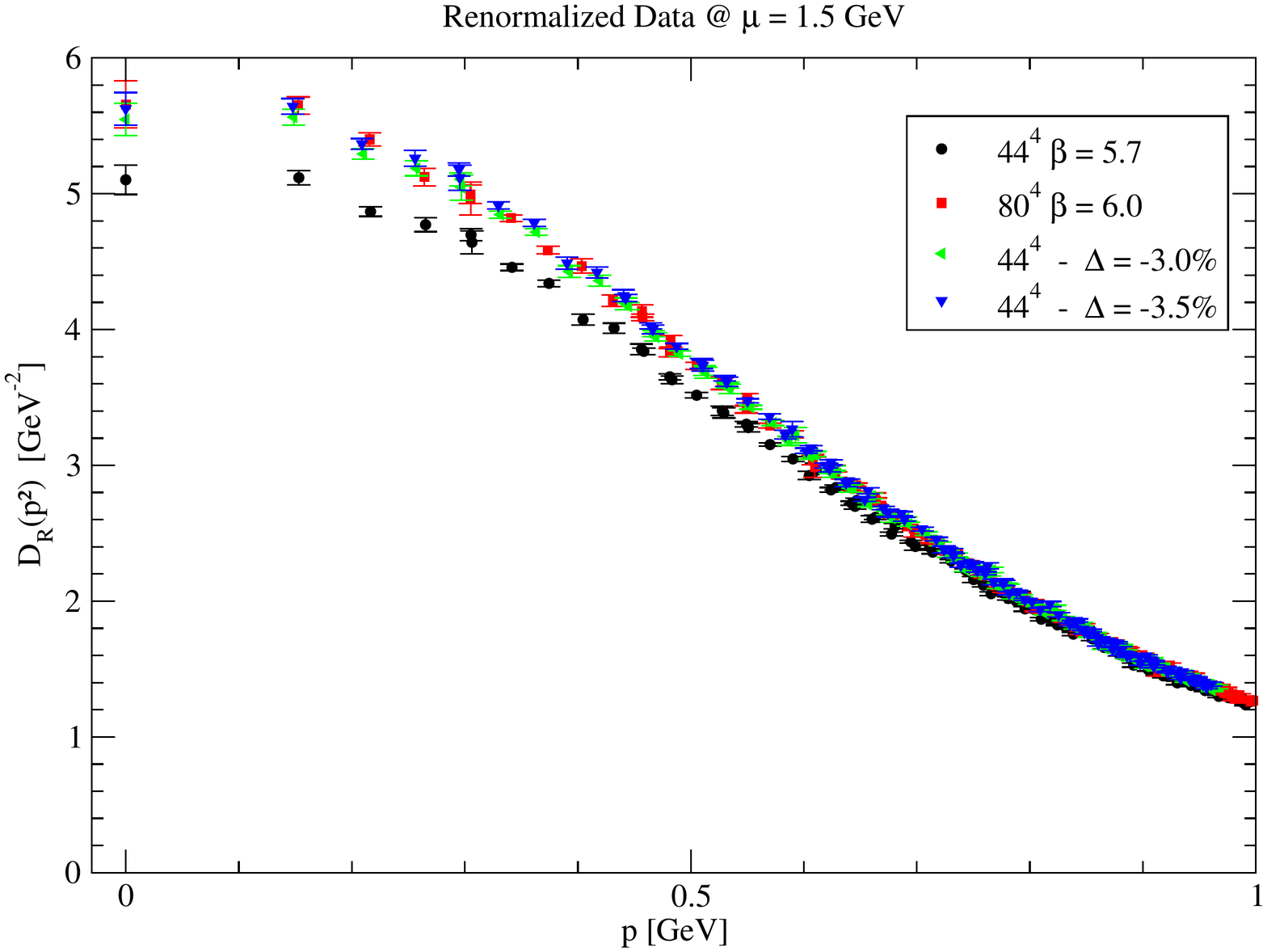} \\
   \includegraphics[width=3in]{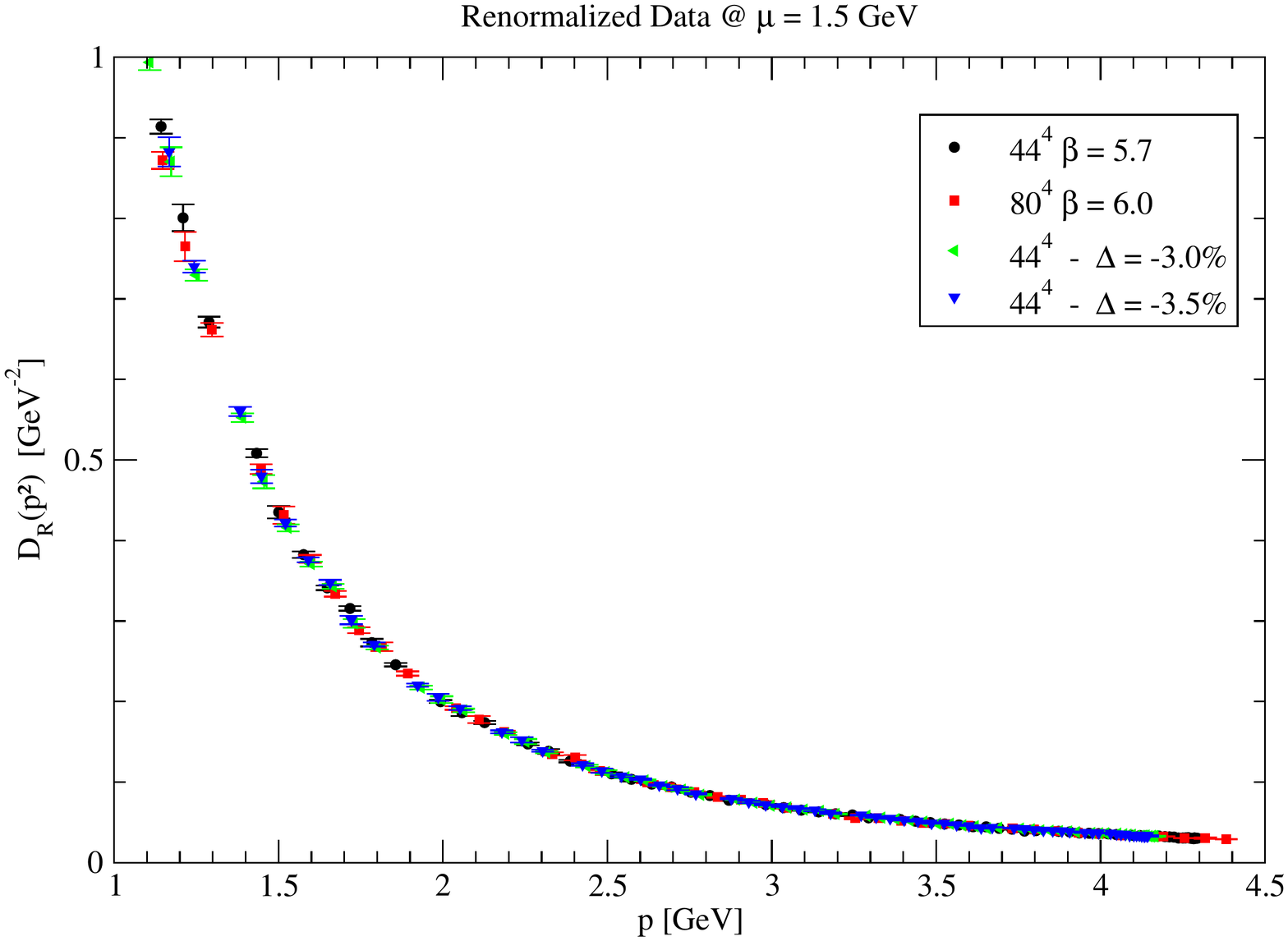}
%
   \includegraphics[width=3in]{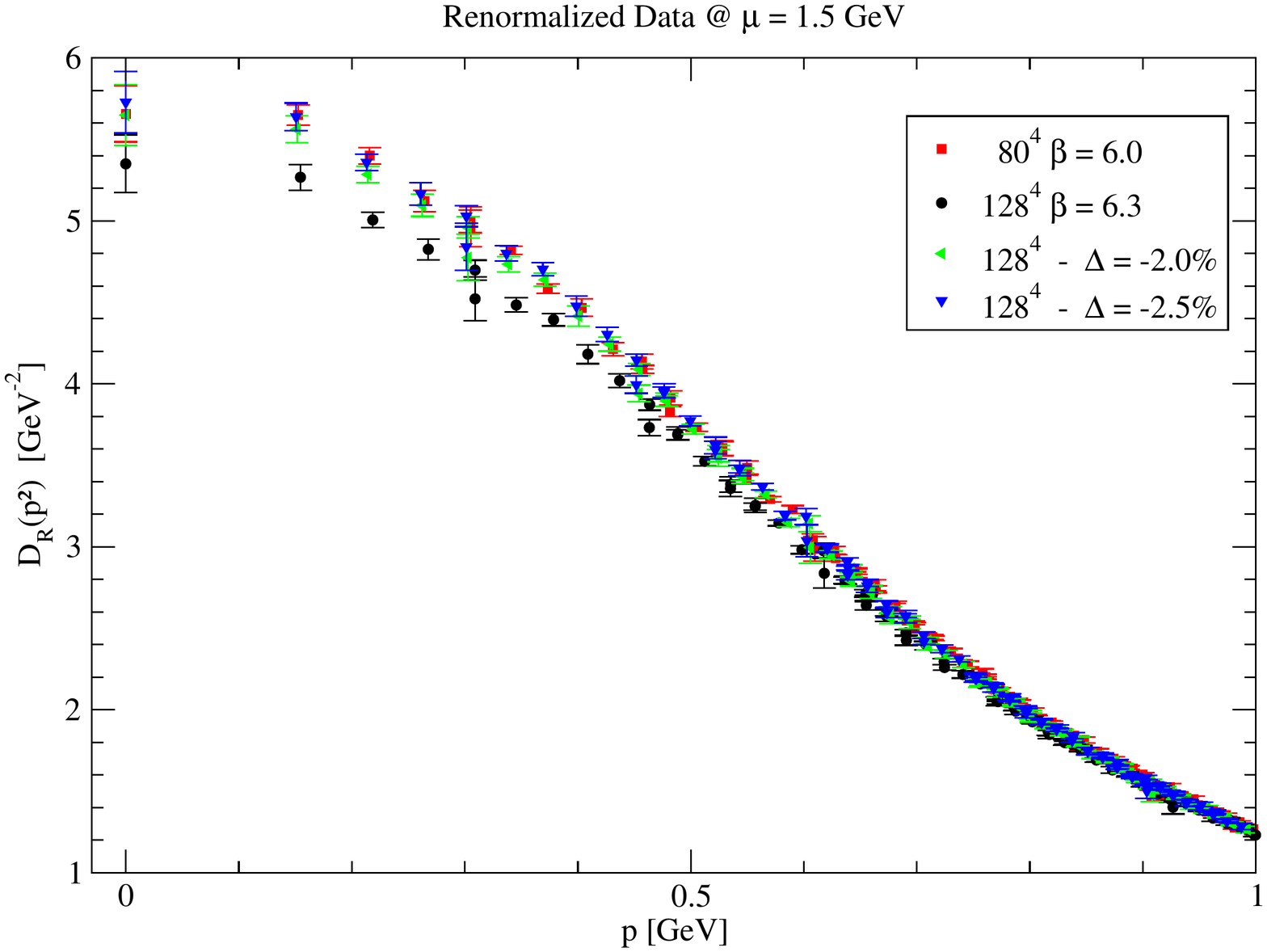}
   \caption{Recalibrated gluon propagator for the $44^4$ and $\beta = 5.7$ data for momenta below 1 GeV (top) and above 1 GeV (middle).
                 The bottom picture shows the $128^4$ and $\beta = 6.3$ recalibrated gluon data for momenta below 1 GeV.}
   \label{fig:props128}
\end{figure}

\begin{figure}[h] 
   \centering
   \includegraphics[width=3in]{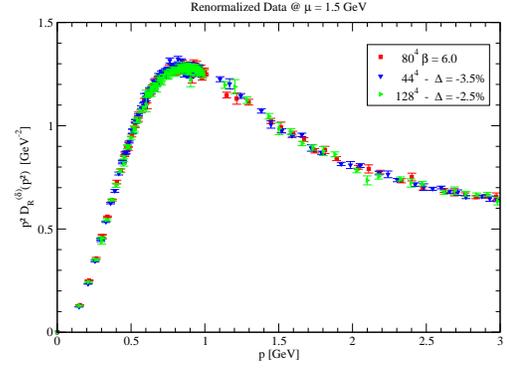}
   \caption{``\textit{Recalibrated}'' gluon dressing function.}
   \label{fig:gluedress}
\end{figure}

\begin{figure}[h] 
   \centering
   \includegraphics[width=3in]{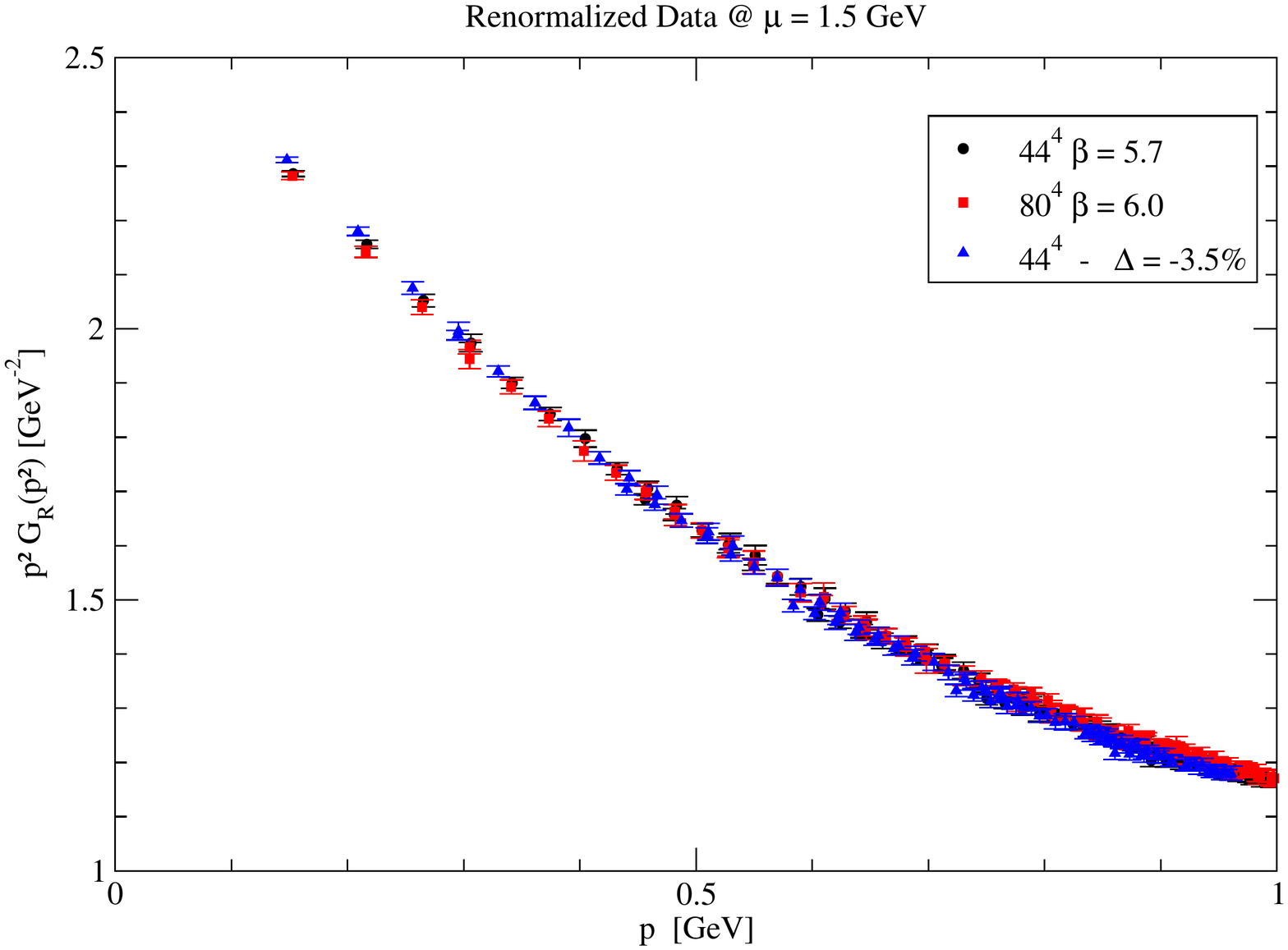} \\
   \includegraphics[width=3in]{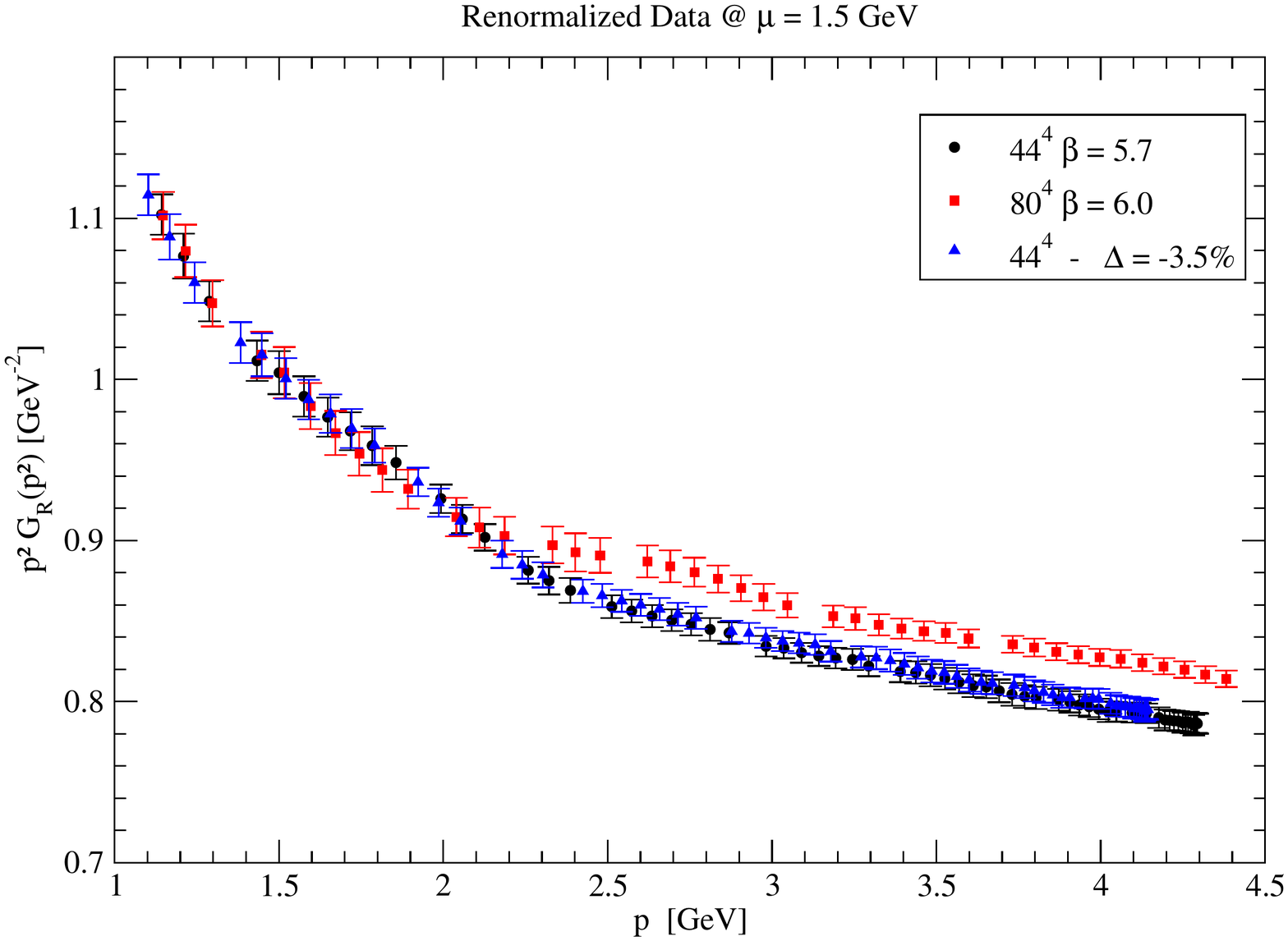}
   \caption{Recalibrated ghost propagator data ($44^4$ and $\beta = 5.7$) for momenta below 1 GeV (top) and above 1 GeV (bottom).}
   \label{fig:ghostprops44}
\end{figure}

The renormalized lattice gluon propagator and ghost dressing function can be seen on Fig.~\ref{fig:props} as a function of tree level improved
momentum $p_\mu = (1/a) \hat{p}_\mu$ and $\hat{p}_\mu = 2 \, \sin ( \pi n_\mu / L )$, with $n_\mu = - L/2, -L/2 + 1, \cdots, 0, \cdots, L/2 -1$ is
the dimensionless lattice momentum.
For the conversion into physical units we used the central value of $a$ reported in Tab. I~\cite{Duarte:2016iko}. 
Clear differences between the various simulations are seen in the infrared gluon data.
In the ghost data, the renormalization at a lower momenta, compared to the choice used
in~\cite{Duarte:2016iko} where $\mu = 4$ GeV, translates into milder differences in the infrared but strong differences in the ultraviolet between the various simulations.
However, in what concerns the dependence with the lattice spacing, the  pattern observed in~\cite{Duarte:2016iko}  
is clearly seen. In particular, the dependence on the lattice spacing for the gluon and ghost  data is opposite, 
with the coarser lattice being below (above) the remaining data for the gluon (ghost) propagator.

Let us follow~\cite{Boucaud:2017a} and allow for a small deviation in the lattice scale $a \rightarrow a^\prime = ( 1 + \delta ) a$.
This rescaling of $a$ translates into a rescaling of the momenta (in physical units)
$p \rightarrow p' = p / (1 + \delta ) = ( 1 + \Delta ) p$; for small corrections $\Delta \sim - \delta$.
The propagators have to rescale
accordingly but, instead, we require the renormalization condition $D_R ( \mu ) = 1 / \mu^2$ to be always fulfilled.
The renormalized propagators, in physical units, computed after the change of scale are named below as ``\textit{recalibrated}''  propagators.
As reference data we take the propagators of the simulation performed using $\beta = 6.0$ and the $80^4$ lattice.


The ``\textit{recalibrated}'' gluon data for the $\beta = 5.7$ and $\beta = 6.3$ can be seen on Fig.~\ref{fig:props128}.
For the first (coarser) lattice data we show the infrared data separately from those with $p \geqslant 1$ GeV. 
Similar curves could be drawn for the (finer lattice) $\beta = 6.3$ data. 
A systematic deviation in the scale setting of the same order of magnitude as the statistical errors
associated to the lattice spacing settles the differences observed on Fig.~\ref{fig:props} both in the infrared and ultraviolet regions. 

The resolution of the differences between the gluon propagator data over the full range of momenta provides a way of setting the
relative values of the lattice spacing either by identifying a particular momenta or by matching the lattice data for different simulations. 
A candidate kinematical 
point
being the maximum of the gluon dressing function, see Fig.\ref{fig:gluedress}.
A naive fit of the data to a Pad\'e approximation given by a $z ( p^2 + m^2_1)/(p^4 + m_2 p^2 + m^4_3)$ in the range $p \in [0.5 , 1.5]$ GeV,
gives $p \sim 0.84$ GeV ($\beta = 5.7$), 0.85 GeV ($\beta = 6.0$) and 0.86 GeV ($\beta = 6.3$) for the maximum of the dressing function. 
An ``exact''  determination of $p_{max}$ demands a detailed and careful analysis. 

In what concerns the ghost propagator, our analysis shows that a `\textit{recalibration}''  
of the lattice spacing does not change significantly the agreement between the different results.
As reported in~\cite{Duarte:2016iko}, the $44^4$ provides the larger $G_R(p^2)$, contrary to the gluon propagator data where 
it provides the lower $D_R(p^2)$. The effects of the lattice spacing on the ghost and gluon propagators seem to point
in opposite directions. Therefore, the procedure of~\cite{Boucaud:2017a} does not seem to be able to improve the agreement between 
simulations simultaneously for both propagators.
On Fig.~\ref{fig:ghostprops44} we report the ``\textit{recalibrated}''  ghost data for the coarser lattice ($\beta = 5.7$).
Similar curves could be reported for 
$\beta = 6.3$ and the larger lattice $128^4$.

In conclusion, our reanalysis of the lattice propagator data published in~\cite{Duarte:2016iko} confirm
that the procedure of~\cite{Boucaud:2017a}  softens the differences between the lattice gluon data for simulations 
with various lattice spacings. However, for the ghost propagator, the recipe does not improve the agreement between the lattice data,
as the deviations are in the opposite direction of the gluon data.
Definitive conclusions concerning the topic discussed here, require a method that provides a good (relative) calibration of the lattice spacing or,
equivalently,  provide a precise lattice measurement of the beta function.
In this sense, a possible method is discussed here, and further work is under development~\cite{Boucaud:2017ab}.  

\begin{acknowledgments}
We thank the Laboratory for Advanced Computing at University of Coimbra 
and the Partnership for Advanced Computing in Europe (PRACE) initiative 
projects COIMBRALATT (DECI-9) and COIMBRALATT2 (DECI-12)
for  the 
computing resources.
The authors acknowledge financial support from F.C.T.  under contract UID/FIS/04564/2016.
P. J. S. acknowledges support by F.C.T. under contract SRFH/BPD/109971/2015. 
\end{acknowledgments}


\bibliographystyle{apsrev4-1}
\bibliography{commentbiblio}{}

\end{document}